**Uma análise bibliométrica do Congresso Nacional de Bibliotecários, Arquivistas e Documentalistas (1985-2012)**

Silvana Roque de Oliveira
Catarina Moreira
José Borbinha
María Ángeles Zulueta Garcia

**RESUMO**

Este artigo constitui a primeira análise bibliométrica das 708 comunicações publicadas pelo Congresso Nacional de Bibliotecários, Arquivistas e Documentalistas entre 1985 e 2012, tendo-se desenvolvido indicadores de produção, de produtividade, de proveniência institucional e análise temática, numa perspectiva quantitativa, relacional e diacrónica. Os seus resultados apresentam um Congresso dinâmico, essencialmente nacional e profissional, com uma forte preponderância da autoria individual, apesar do recente crescimento da taxa de colaboração. Na sua abordagem temática, é dado realce aos serviços públicos de informação, com maior destaque para o mundo das bibliotecas, mantendo também importância as reflexões sobre a formação profissional e académica na área da Ciência da Informação, bem como o acompanhamento dos mais recentes desenvolvimentos tecnológicos.

**PALAVRAS-CHAVE:**

Congresso Nacional de Bibliotecários, Arquivistas e Documentalistas

APBAD

1985-2012

Análise bibliométrica

Ciência da Informação

Portugal

**Uma análise bibliométrica do *Congresso Nacional de Bibliotecários, Arquivistas e Documentalistas* (1985-2012)**

Silvana Roque de Oliveira
Catarina Moreira
José Borbinha
María Ángeles Zulueta García

## Introdução

Este artigo apresenta uma análise bibliométrica das onze edições do Congresso Nacional de Bibliotecários, Arquivistas e Documentalistas - doravante referido como o Congresso - a saber: Porto, 1985; Coimbra, 1987; Lisboa, 1990; Braga, 1992; Lisboa, 1994; Aveiro, 1998; Porto, 2001 Estoril, 2004; Ponta Delgada, 2007; Guimarães, 2010; e Lisboa, 2012.

Como resultado de uma abordagem quantitativa, relacional e diacrónica da produção, autoria, proveniência institucional e temáticas abordadas, fornecem-se dados que podem ajudar a melhor entender o papel do Congresso na promoção da Ciência da Informação em Portugal.

## Revisão da literatura

Consideramos que o facto da Ciência da Informação ser, ainda hoje, encarada como uma ciência emergente no panorama académico português (CALIXTO, 2008), aliado à fraca presença internacional e à falta de interesse que a Bibliometria continua a merecer em Portugal, tem contribuído para a escassez de estudos quantitativos na área.

O caso português começou por ser observado de forma comparativa no contexto da produção ibero-americana (MOYA-ANEGÓN; HERRERO-SOLANA, 2002, e HERRERO-SOLANA; LIBERATORE, 2008), em dois estudos realizados com dados recolhidos na *Web of Knowledge* (http://wokinfo.com/), o primeiro entre 1991-2000 e o segundo prolongando a análise até 2005. O retrato aí traçado apresenta a contribuição portuguesa com valores genericamente baixos, exceptuando o de autor mais produtivo para a década de noventa, onde Ana Maria Ramalho Correia, então investigadora do INETI (actual LNEG), atinge o primeiro lugar, *ex aequo* com Isabel Gómez do CINDOC (Espanha). De toda a maneira, como sublinham de forma assertiva Moya-Anegón e Herrero-Solana (2002), a produção local da região ficou por estudar, pois só uma muito pequena parcela tinha visibilidade naquela base de dados, maioritariamente a produção editada no mundo anglo-saxónico. Do mesmo modo, num artigo exclusivamente dedicado à produção portuguesa na área (OLMEDA

GÓMEZ; PERIANES-RODRÍGUEZ; OVALLE-PERANDONES, 2008) voltou-se a recorrer à *Web of Knowledge*, ficando mais uma vez por caracterizar a produção editada em Portugal.

No contexto nacional, a primeira análise ficou a dever-se a Laura Cerqueira e Armando Malheiro da Silva (2007), sobre os dez anos da revista *Páginas a&b: arquivos & bibliotecas*, onde, para alguns indicadores, se estabeleceram comparações com os *Cadernos BAD*. Mais recentemente foi publicado um levantamento bastante exaustivo da produção portuguesa em acesso aberto em Ciência da Informação (RIBEIRO; PINTO, 2009) e um estudo exploratório, de cariz mais qualitativo, sobre a produção portuguesa na área temática da Organização e Representação do Conhecimento (RIBEIRO, 2012).

As comunicações publicadas em actas de congressos têm vindo a merecer um crescente reconhecimento do seu papel singular na comunicação científica, não só por oferecerem boas condições para se partilharem e debaterem as mais novas tendências de cada ramo do conhecimento, como por constituírem excelentes oportunidades para os investigadores alargarem as suas redes de contactos e potenciarem futuras colaborações (GONZÁLEZ-ALBO; BORDONS, 2011). Já no que toca a sua posterior consulta entre a comunidade científica, há estudos que apontam para um decréscimo muito acentuado das citações de actas face a outras tipologias documentais, como os artigos de revista, excepção feita para áreas como a Engenharia Informática (LISÉE; LARIVIÈRE; ARCHAMBAULT, 2008). Não obstante, no caso de áreas com um acentuado perfil técnico e profissional como é a nossa, continua também a considerar-se que a análise dos congressos mantém grande pertinência (LÓPEZ-CÓZAR, 2002, e GLÄNZEL; SCHLEMMER; SCHUBERT, 2006). Surgiu assim a motivação para este estudo sobre os Congressos.

**Material e métodos**

Visto não estar acessível de forma estruturada, a informação foi directamente recolhida das versões integrais dos volumes de actas e registada numa base de dados criada para o efeito (em Microsoft SQL). Como unidade de análise elegemos as comunicações no seu sentido mais estrito, limitando-as às contribuições apresentadas no âmbito das sessões de trabalho e publicadas na sua versão completa. Assim, ficaram de fora os discursos, resumos de comunicações e as sínteses de mesas-redondas, *posters* ou apresentações orais.

Como resultado foram identificadas, nas onze edições dos Congressos, 708 comunicações, 744 autores e 437 instituições. Para o efeito foram também compulsadas fontes secundárias disponíveis na Internet (currículos e outras informações referentes aos autores e instituições à data das comunicações em causa). Assim, elementos como a origem institucional e geográfica, ou o desenvolvimento de algumas abreviaturas de nomes próprios foram explicitados fora do contexto bibliográfico das actas, colmatando lacunas criadas por deficiências formais ainda persistentes na edição dos textos, maioritariamente até ao oitavo congresso, em 2001. A partir dessa data a fixação de um estilo de edição, acompanhado do acesso digital, vieram normalizar os textos, facilitando o tratamento bibliométrico.

Com base neste universo, desenvolvemos uma abordagem quantitativa com um enfoque diacrónico para algumas das análises. O intervalo temporal dos Congressos (1985-2012) foi dividido em dois momentos: 1985-1998 (6 edições) e 2001-2012 (5 edições). Para esta divisão, baseámo-nos numa repartição o mais equitativa possível num conjunto ímpar de edições, corroborada por uma feliz coincidência conjuntural – o facto de 2001 ser o ano do lançamento da primeira Licenciatura em Ciência da Informação, na Universidade do Porto, em ruptura com o anterior modelo formativo (SILVA, 2002) – para além de ser o início de uma década particularmente profícua na oferta de formação académica na área (PINTO, 2008). A análise recorreu a indicadores unidimensionais (indicadores de produção, de produtividade, de proveniência institucional e análise temática através de uma classificação aplicada manualmente) e multidimensionais (co-autoria nominal, análise de colaboração e análise temática complementar, automatizada, através da análise da co-ocorrência de palavras-chave). Para maior comodidade na leitura, as opções metodológicas específicas a cada um dos indicadores serão apresentadas no subcapítulo que lhe corresponde.

## Análise e Discussão dos Resultados

### Análise da produção

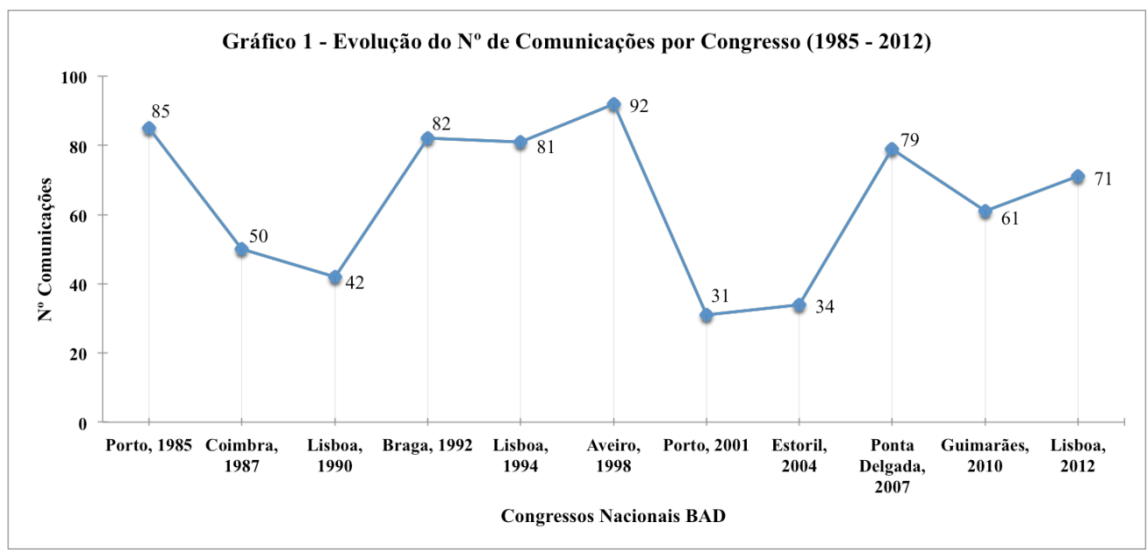

Do **Gráfico 1**, começamos por destacar as alterações na regularidade de realização do Congresso, oscilando entre os dois e os quatro anos de intervalo entre edições, tendo sido o VI Congresso, em Aveiro, o de maior nível de participação, com 92 comunicações. A primeira edição (Porto, 1985) apresenta o segundo valor mais elevado de comunicações (85), prova do bom acolhimento desta nova iniciativa da BAD. Recorda-se que em Portugal, entre 1982 e 1985, estava a ter lugar a reestruturação dos cursos de nível superior na área, inalterada durante cerca de 50 anos, tendo-se fundado três Cursos de Especialização em Ciências Documentais (CECD), nas universidades de Coimbra (1982), Lisboa (1983) e Porto (1985), sinal da vitalidade da disciplina, mesmo que em moldes ainda marcadamente profissionalizantes (PINTO, 2008). No entanto, a procura que se fez sentir no I Congresso desceu acentuadamente nas segunda e terceira edições, apesar de se realizarem precisamente nas outras duas cidades que tinham lançado o novo modelo de estudos pós-graduados, demonstrando que os novos impulsos dados à formação académica não tiveram impacto na adesão aos Congressos. No início da década de 2000 verificou-se uma descida muito acentuada, para cerca de um terço, durante as duas primeiras sessões (Porto, 2001 e Estoril, 2004), voltando a subir muito significativamente na edição realizada em 2007, em Ponta Delgada. Nas duas últimas edições, oscilou-se entre uma descida para 61 comunicações, em 2010, e uma subida para 71 em 2012.

**Análise da Autoria**

Para a análise da autoria pessoal e para a institucional foi necessária uma normalização de nomes, a fim de garantir uma contagem consistente das frequências, isenta de distorções causadas pela sua alteração ao longo do tempo. Tal situação foi mais comum entre os autores do sexo feminino, por mudanças no estado civil, e entre as instituições públicas, por constantes reestruturações, tendo-se optado pela denominação mais recente, ou, no caso de alguns ministérios, pela redução do nome aos seus elementos mais constantes.

**Autoria por Género**

A distribuição dos autores dos Congressos por género é o indicador mais estável ao longo do tempo, revelando uma predominância do sexo feminino (64,2% das comunicações entre 1985-1998, e 63,7% entre 2001-2012). Mais uma vez, se atendermos aos autores mais produtivos ao longo das 11 sessões (**Tabela 1**), voltamos a encontrar uma maioria feminina, o que contraria a tendência de uma maior produtividade masculina na área da Ciência da Informação, quando nos encontramos em contexto exclusivamente académico (ARQUERO AVILÉS, 2001). Sendo um dado muito sensível a diversos aspectos circunstanciais, esta aparente contradição leva-nos a crer que será necessário desenvolver outros indicadores para melhor se entenderem as suas razões.

| Tabela 1. Autores com 4 ou mais Comunicações (1985-2012) | | | |
|---|---|---|---|
| **Género** | **Feminino** | **Masculino** | **Total** |
| Número de Autores | 33 | 20 | 53 |
| Distribuição relativa | 64,5% | 35,5% | |
| Número de Comunicações | 205 | 113 | 318 |
| Distribuição relativa | 62,3% | 37,7% | |

**Análise da produtividade dos autores**

Para a análise da produtividade de autores e instituições, seguimos o método do cálculo completo (MALTRAS, 2003), atribuindo a todos os co-autores o peso de uma unidade por comunicação, acreditando que a relação do esforço despendido numa comunicação individual ou em parceria é demasiado complexa para se distinguir pela simples proporção directa fraccionada, valorizando-se assim positivamente os autores em colaboração.

Na **Tabela 2**, listamos os autores com 4 ou mais comunicações nas onze edições. Além do já comentado quanto ao género, sublinhamos um denominador comum aos primeiros 7 nomes, com 10 ou mais comunicações: o facto de que todos têm vindo a apostar numa formação académica de topo na área específica da Ciência da Informação ou dos Sistemas de Informação. Tal continua a aplicar-se a outros autores da mesma lista, de forma intercalada, o que poderia remeter para um perfil essencialmente académico dos autores mais produtivos. No entanto, apesar de muitos destes autores acumularem a docência com a actividade bibliotecária ou arquivística, esta última ainda permanece a principal ocupação - excepção para José Borbinha - o que nos remete para o paradigma essencialmente profissional da área, mesmo quando muitos dos seus actores já reúnem as competências exigidas pelo universo académico.

**Tabela 2. Lista dos Autores com 4 ou mais Comunicações (1985-2012)**

| Autores | Nº Comunicações |
|---|---|
| 1. CALIXTO, José António | 16 |
| 2. OCHÔA, Paula | 16 |
| 3. CAMPOS, Fernanda Maria Guedes | 14 |
| 4. AMANTE, Maria João | 14 |
| 5. PINTO, Leonor Gaspar | 13 |
| 6. BORBINHA, José Luís | 10 |
| 7. CORDEIRO, Maria Inês | 10 |
| 8. RODRIGUES, Eloy | 8 |
| 9. CABRAL, Maria Luísa | 7 |
| 10. SILVA, Armando Malheiro da | 7 |
| 11. SILVA, Vera Maria da | 6 |
| 12. COSTA, Maria Teresa | 6 |
| 13. FERREIRA, Maria Fernanda Casaca | 6 |
| 14. FIGUEIREDO, Fernanda Eunice | 6 |
| 15. FREIRE, Nuno | 6 |
| 16. GALVÃO, Rosa Maria Tavares | 6 |
| 17. RIBEIRO, Fernanda | 6 |
| 18. LIMA, Maria João da Silva Pires de | 6 |
| 19. MARTINS, Ana Bela de Jesus | 6 |
| 20. GRAÇA, Almerinda | 5 |
| 21. RAMALHO, José Carlos | 5 |
| 22. OLIVEIRA, Margarida P. | 5 |
| 23. PAIVA, Lucília | 5 |
| 24. FERREIRA, Miguel | 5 |
| 25. FARIA, Isabel | 5 |
| 26. BORGES, Leonor Galvão | 5 |
| 27. CARDOSO, Armindo R. | 5 |
| 28. BARRETO, Adalberto | 5 |
| 29. BARRULAS, Maria Joaquina | 5 |
| 30. AMADO, João Paulo | 5 |
| 31. ANTÓNIO, Rafael | 5 |
| 32. SANTOS, Maria Luísa Nunes dos | 5 |
| 33. SARAIVA, Ricardo | 4 |
| 34. SILVA, Maria Albertina Melo Marcos da | 4 |
| 35. SOUSA, José Manuel Mota de | 4 |
| 36. ALVES, Luísa M. P. A. | 4 |
| 37. CARDOSO, João Carlos | 4 |
| 38. CARVALHO, José | 4 |
| 39. EIRAS, Bruno Duarte | 4 |
| 40. CORREIA, Ana Maria Ramalho | 4 |
| 41. CORREIA, Zita | 4 |
| 42. GABRIEL, Graça da Conceição Filipe | 4 |
| 43. GONÇALVES, João da Silva | 4 |
| 44. MOURA, Maria José | 4 |
| 45. PRÍNCIPE, Pedro | 4 |
| 46. RAFAEL, Gina Guedes | 4 |
| 47. REI, Maria da Luz Nogueira | 4 |
| 48. LÓPEZ GÓMEZ, Pedro | 4 |
| 49. LOURENÇO, Maria Alexandra | 4 |
| 50. LEAL, Filipe | 4 |
| 51. MARTINS, Lígia Maria de Azevedo | 4 |
| 52. MARIANO, Emília Henriques Gouveia da Silva | 4 |
| 53. MELO, Luiza Baptista | 4 |

**Análise da Co-Autoria**

A co-autoria tem vindo a ser crescentemente valorizada na comunidade científica, por se entender que a responsabilidade partilhada constitui a primeira forma de controlo de qualidade dos resultados (ROMÁN ROMÁN, 2001). Estudos têm indicado serem as publicações em co-autoria reconhecidas com maior autoridade, pelo maior número de citações recebidas (GLÄNZEL, 2002, BEAVER, 2004).

A evolução da média do número de autores assinantes por comunicação (**Tabela 3**) mostra um crescimento – como curiosidade, para a *Revista Española de*

*Documentación Científica* foi calculado um índice de 2,4, entre 1997-2005 (JÍMENEZ HIDALGO, 2007).

| **Tabela 3.** Evolução do Índice de Co-Autoria ||
|---|---|
| **Intervalos Cronológicos** | **Índice de Co-Autoria** |
| Congressos 1985-1998 | 1,5 |
| Congressos 2001-2012 | 2 |

Mais em detalhe, a relação do número de comunicações por número de autores (**Tabela 4**) reforça a imagem de uma comunidade ainda bastante individualista, típico das Ciências Sociais (CHINCHILLA RODRÍGUEZ; MOYA ANÉGON, 2007), apesar do expressivo aumento da taxa de colaboração entre os dois intervalos temporais.

| **Tabela 4. Evolução do Nº de Autores por Comunicação** |||
|---|---|---|
| **Nº Autores/ Comunicação** | **Congressos 1985-1998** | **Congressos 2001-2012** |
| 1 Autor | 70% | 46% |
| 2 Autores | 17% | 31% |
| 3 Autores | 8% | 15% |
| ≥ 4 Autores | 5% | 8% |
| **Taxa de Colaboração** (% Comunicações ≥ 2 Autores) | **30%** | **54%** |

**Análise de redes de co-autoria**

As visualizações das redes de co-autoria foram criadas pelo algoritmo *Kamada-Kawai* no programa Pajek (http://pajek.imfm.si), segundo os dados extraídos da base de dados. Na impossibilidade de obtermos grafos para impressão legível com a totalidade dos autores que participaram nos Congressos, limitámo-nos a reduzir o universo aos autores que assinaram comunicações com autores de outras instituições, não sendo necessário recorrer a algoritmos de selecção de subgrafos, método já utilizado num trabalho anterior (OLIVEIRA et al., s.d.).

Um grafo pode ser definido como um conjunto de nós interligados entre si através de um conjunto de ligações (LEMIEUX; OUIMET, 2004). O tamanho de cada nó, que representa cada autor, é proporcional ao número de documentos em que esse autor ocorre; a grossura das ligações transmite o peso que corresponde ao número de vezes que uma co-autoria ocorre. Ilustramos a evolução da co-autoria em dois grafos, correspondentes aos dois intervalos de tempo usados.

A primeira rede (1985-1998) (**Grafo 1**) mostra uma comunidade ainda dispersa, com pólos como pequenas ilhas isoladas, com ligações muito finas. Como grupos, sobressaem as redes em torno de Margarida P. Oliveira e Lucília Paiva, e em torno de Ana Maria Ramalho Correia. Pertencendo à primeira rede, Fernanda Maria Guedes Campos sobressai como a autora com mais comunicações em colaboração, ilustrado pelo tamanho do respectivo nó.

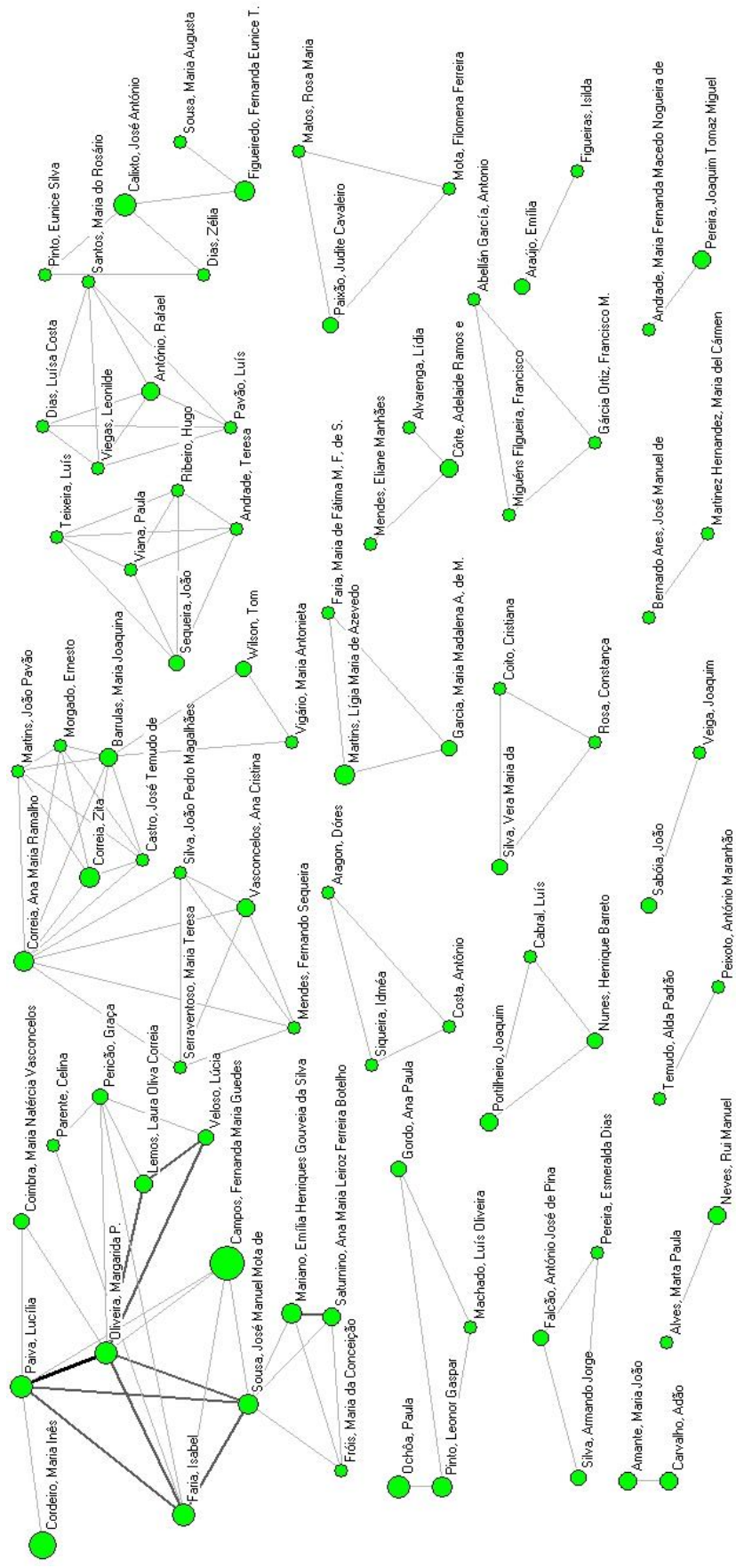

**Grafo 1**. Co-autores com colaboração de tipo nacional e internacional (Congressos 1985-1998) Pajek 3.11

**Grafo 2. Co-autores com colaboração de tipo nacional e internacional (Congressos 2001-2012) Pajek 3.11**

No **Grafo 2**, entre 2001-2012, multiplicam-se as relações de co-autoria, destacando-se as redes onde participam José António Calixto, José Luís Borbinha, Maria João Amante, Maria Teresa Costa e Paula Ochôa. De assinalar as fortes relações de co-autoria entre José Luís Borbinha, Nuno Freire e Hugo Manguinhas, bem como entre Paula Ochôa e Leonor Gaspar Pinto. Com ligações menos intensas, mas de dimensões também assinaláveis, atente-se nas redes em torno de José Carlos Ramalho (10 autores) e Luiza Melo Baptista (9 autores).

### Análise da afiliação institucional dos autores

A proveniência institucional foi recuperada no primeiro endereço dado por cada autor, tendo-se identificado o departamento, o organismo em que se enquadrava, o distrito e o país, atribuindo-se por fim uma classificação tipológica a cada instituição. Nesta classificação, estabeleceram-se oito classes principais, tal como constam da **Tabela 5**. Antes de analisarmos os dados, é conveniente explicitar os critérios de classificação adoptados para algumas das classes. Partindo-se das propostas de Maria del Rosario Arquero Avilés (2001) fizeram-se adaptações ao contexto português. Dentro dos "Organismos de Gestão da Administração" estão todos os departamentos de administração estatal, dos municipais aos internacionais, mesmo os ligados à área BAD. As universidades reagrupam tanto as faculdades e centros de investigação a elas ligados, como os institutos politécnicos ou os laboratórios de investigação científica, privilegiando-se assim como critério a promoção directa das actividades académicas e científicas de nível superior. As "Associações" congregam a própria BAD, bem como algumas congéneres nacionais e internacionais. As "empresas" reúnem tanto as privadas como as públicas. Finalmente, em "Outros", classificámos os hospitais, as escolas do Ensino Básico e Secundária, as fundações e entidades transnacionais.

| Tabela 5. Evolução da distribuição relativa das Comunicações por Tipologia Institucional (1985-2012) | | |
|---|---|---|
| **Tipologias Institucionais** | **Comunicações 1985-1998** | **Comunicações 2001-2012** |
| Bibliotecas | 38,6% | 45,0% |
| Universidades & Investigação | 18,9% | 31,4% |
| Organismos de Gestão da Administração | 14,1% | 10,6% |
| Arquivos | 12,4% | 5,1% |
| Associações | 5% | 1,1% |
| Outros | 4,2% | 2,2% |
| Empresas | 3,4% | 2,7% |
| S/ Afiliação | 3,4% | 1,9% |

Pela leitura da **Tabela 5**, pode afirmar-se a manifesta preponderância das "Bibliotecas" ao longo dos anos. Por sua vez, as "Universidades & Investigação" protagonizam o maior aumento, vindo quase a duplicar a sua presença. Em sentido contrário, os "Arquivos", não só ocupam um tímido quarto lugar, como protagonizam a maior queda na participação nos Congressos. É ainda de notar a presença residual das "Empresas".

**Análise da produtividade institucional**

De entre as instituições mais produtivas de 1985 a 2012 (**Tabela 6**) destaca-se a Biblioteca Nacional de Portugal, com 83 comunicações em que pelo menos um dos autores a representa, seguida de longe por um grupo liderado pelas Bibliotecas Municipais de Oeiras. Mais uma vez se confirma a predominância das "Bibliotecas", ocupando os cinco lugares cimeiros, surgindo a primeira instituição arquivística, os Arquivos Nacionais, apenas na décima posição. Note-se, finalmente, uma clara supremacia das instituições públicas, apenas contrariada por excepções pontuais.

**Tabela 6. Lista das Instituições com 4 ou mais Comunicações (1985-2012)**

| Instituições | Nº Comunicações |
|---|---|
| 1. Biblioteca Nacional de Portugal | 83 |
| 2. Bibliotecas Municipais de Oeiras | 14 |
| 3. Serviços de Documentação e Informação Univ. Minho | 13 |
| 4. Serviços de Documentação e Informação Univ. Aveiro | 12 |
| 5. Biblioteca Geral Univ. Coimbra | 12 |
| 6. CITI LNEG | 12 |
| 7. Bibliotecas Municipais de Lisboa | 12 |
| 8. Faculdade de Letras Univ. Porto | 10 |
| 9. Biblioteca de Arte (Fundação Calouste Gulbenkian) | 10 |
| 10. Arquivos Nacionais/Torre do Tombo | 9 |
| 11. Biblioteca do ISCTE | 9 |
| 12. Biblioteca Pública de Évora | 9 |
| 13. Divisão de Bibliotecas e Documentação CM Lisboa | 8 |
| 14. Departamento de Informática Univ. Minho | 7 |
| 15. CIDEHUS Univ. Évora | 7 |
| 16. Archivo del Reino Galícia | 7 |
| 17. Arquivo Distrital de Braga Univ. Minho | 6 |
| 18. Arquivo Distrital do Porto | 6 |
| 19. B-On FCCN | 6 |
| 20. Biblioteca Municipal do Seixal | 6 |
| 21. Fac. Comunicación y Documentación Univ. Granada | 6 |
| 22. Serv Doc Info Fac Eng UP | 6 |
| 23. Serv. Informação Científica Técnica Minist. Seg. Social | 6 |
| 24. INESC-ID – Grupo Sistemas da Informação UTL | 5 |
| 25. Divisão Gestão Arquivos CM Lisboa | 5 |
| 26. DGLB | 5 |
| 27. Departamento Biblioteconomia Univ. Brasília | 5 |
| 28. Biblioteca Fac. Farmácia Univ. Coimbra | 5 |
| 29. Biblioteca Fac. Ciências Tecnologias Univ. Coimbra | 4 |
| 30. Arquivo Histórico Municipal Porto | 4 |
| 31. British Library | 4 |
| 32. CDI ESE IP Coimbra | 4 |
| 33. CDI Secretaria-Geral Ministério Educação | 4 |
| 34. Departamento Biblioteconomia e Documentação UNESP | 4 |
| 35. Bibliothèque nationale de France | 4 |
| 36. International Council of Archives (ICA) | 4 |
| 37. Departamento História Univ. Portucalense | 4 |
| 38. Departament Infomation Studies Univ. Sheffield | 4 |
| 39. Escola Biblioteconomia Univ. Fed. Minas Gerais | 4 |
| 40. Fac. Ciências Sociais Humanas Univ. Nova Lisboa | 4 |
| 41. KEEP SOLUTIONS, LDA | 4 |
| 42. Laboratório José de Figueiredo IGESPAR | 4 |

**Análise da Colaboração institucional**

Na **Tabela 7**, pode observar-se a evolução da autoria distribuída por quatro tipos de colaboração institucional: intra-departamental (só autores do mesmo departamento); intra-institucional (autores da mesma instituição, mas com pelo menos um de um departamento diferente); nacional (autores de pelo menos duas instituições distintas); internacional (pelo menos um autor de um segundo país). Tenha-se presente que os dados recolhidos se reportam exclusivamente ao universo das comunicações com colaboração (30% e 54% das comunicações, para os dois intervalos cronológicos adoptados). Para o período total de 27 anos, é persistente o peso da colaboração intra-departamental, evidenciando uma replicação da proximidade laboral no perfil de colaboração da maioria dos autores. A maior mudança ocorre ao nível da colaboração internacional.

| Tabela 7. Evolução do Tipo de Colaboração nas Comunicações (1985-2012) | | |
|---|---|---|
| **Tipologia de Colaboração** | **1985-1998** | **2001-2012** |
| Intra-Departamental | 65% | 53% |
| Intra-Institucional | 8% | 4% |
| Nacional | 26% | 34% |
| Internacional | 0,7% | 9% |

Recorrendo às mesmas ferramentas utilizadas para a análise de redes de co-autoria nominal, também seria interessante apresentar a evolução das relações de co-autoria institucional. A informação estruturada na base de dados permite essa análise, mas, infelizmente, devido à sua complexidade, não foi possível terminá-la a tempo de publicação neste artigo (essa complexidade deve-se ao grande aumento do número de colaborações fora da mesma instituição, como se pode ver comparando os períodos na **Tabela 7,** o que obrigaria a outras estratégias para a visualização da informação).

**Análise temática**

A análise temática foi desenvolvida com recurso a dois métodos complementares, já seleccionados num trabalho anterior (OLIVEIRA et al., s.d): a classificação temática manual, que conseguimos aplicar a todo o período estudado, e a co-ocorrência de palavras-chave, só possível entre 2001-2012, anos em que as actas passam a incluir palavras-chave escolhidas pelos autores.

**Classificação temática com vocabulário controlado**

Desenvolvemos a classificação manual dos 708 documentos, adoptando o esquema proposto por Järvelin e Vakkari (1990 e 1993), utilizado por López-Cózar (2002) e retomado por Rochester e Vakkari (2003). Para algumas classes, muito conotadas com as bibliotecas, optou-se por designações mais abrangentes, capazes de integrarem a Arquivística. A lista final apresenta os seguintes termos: Profissão;

Evolução dos Serviços; Evolução dos Documentos & Edição; Serviços de Informação; Organização e Recuperação da Informação; Busca de Informação; Comunicação Científica; Outros Tópicos; e Outras Disciplinas. A cada comunicação foi atribuída uma classe. Uma vez que não procedemos a nenhuma distinção metodológica das comunicações, nem sequer as dividimos entre contribuições de carácter mais geral, técnico ou científico, em cada classe estão agrupados trabalhos que abordam as temáticas com diferentes graus de cientificidade.

Pela **Tabela 8** podemos confirmar o lugar central que ocupa o tema dos "Serviços de informação" - votado a todos os aspectos que se prendem mais directamente com a gestão e planeamento dos serviços, desde a constituição das colecções, as actividades pensadas para o público, a avaliação do desempenho, até à concepção dos próprios edifícios -, com cerca de 40% de comunicações ao longo de todas as edições. Acima dos 10% surgem só mais duas temáticas: "Organização e Recuperação da Informação" - com as questões centrais da descrição e recuperação da informação, contemplando também a preservação física e digital – que desce de 25% para 13%, a partir de 2001; a "Busca da Informação" - com as diversas facetas do comportamento informacional, bem como os canais de acesso à informação -, que, em compensação, sobe dos 7,7% para os 15,6%. Uma nota final para as "Outras disciplinas", onde classificámos textos como estando ligados aos Sistemas de Informação e às Ciências da Educação, residuais mas relevantes a partir do Congresso de 2001.

| Tabela 8. Evolução temática das comunicações nos Congressos (1985-2012) | | |
|---|---|---|
| **Temas** | **1985-1998** | **2001-2012** |
| Profissão | 3,9% | 5,1% |
| Evolução dos Serviços | 3,9% | 3,9% |
| Evolução dos Documentos & Edição | 2,8% | 1,4% |
| Formação em CI | 9% | 8% |
| Serviços de Informação | 39,6% | 41,7% |
| Organização & Recuperação da Informação | 25% | 13,4% |
| Busca de Informação | 7,7% | 15,6% |
| Comunicação Científica | 0,2% | 2,5% |
| Outros Temas | 7,9% | 5,1% |
| Outras Disciplinas | 0% | 3,3% |

**Análise temática com co-ocorrência de palavras-chave**

A análise de co-ocorrência de palavras tem demonstrado ser eficiente a identificar tendências temáticas a partir das relações estabelecidas entre pares de palavras (HE, 1999), recorrendo a formas automatizadas de observação, no quadro conceptual da análise de redes sociais (CALLON et al., 1983, e YI; CHOI, 2012). Numa comparação entre a classificação por meio de *thesauri* especializados e a co-ocorrência de palavras-chave, esta última revelou-se muito eficaz para captar o dinamismo da evolução temática da área, dada a variância da linguagem natural, manifestando-se assim uma interessante complementaridade entre as duas abordagens (DING; GHOWDHURY; FOO, 2000). Têm surgido estudos que aplicam a análise de co-ocorrência de palavras à caracterização temática da Ciência da Informação, com base em diferentes tipologias documentais, recorrendo às palavras-chave ou aos títulos das publicações (MILOJEVIC et al. 2011; LIU et al., 2012; ZONG, et al., 2013).

À semelhança das anteriores análises de redes, também aqui reduzimos previamente a nossa amostra, seleccionando os descritores com pelo menos quatro frequências absolutas, o que resultou em 10 descritores entre 2001-2006 e 52 descritores entre 2007-2012, de um total de 596 descritores.

Da análise conjunta dos **Grafos 3** e **4**, fica bem patente a maior diversidade temática das últimas três edições do Congresso (2007-2012), que apresentam uma rede mais densa, onde surgem temas novos, apenas observáveis por estarmos em contexto de linguagem natural. Em 2001-2006 (**Grafo 3**), surpreendemos apenas dois conjuntos temáticos fundamentais: a Formação, aliado à Ciência da Informação e às suas áreas aplicadas da Arquivística e da Biblioteconomia, bem como aos profissionais, ao lado do universo digital, também ele associado à necessidade de formação. Estes dois temas vão continuar a merecer destaque no segundo intervalo (**Grafo 4**), mas os nós mais proeminentes passarão a ser as Bibliotecas Públicas, o descritor com maior relação com o descritor Portugal, a mostrar como a RNBP é vista como um lugar privilegiado de observação do país no que toca os sistemas de informação. Para além deste tema central, Portugal está também associado às Bibliotecas Universitárias, novo nó de onde irradiam densas relações com as actuais questões da Literacia da Informação. Num terceiro nível, o descritor Portugal liga-se aos Recursos Electrónicos e ao movimento do *Open Access*. Outros dois pólos temáticos destacados são a Avaliação e os Indicadores de Desempenho, essencialmente desenvolvidos no quadro das Bibliotecas, e a questão da Interoperabilidade e dos Metatados, onde surgem grandes projectos internacionais como a *Europeana*, e o universo dos Arquivos, cuja presença ainda se pauta por uma certa timidez.

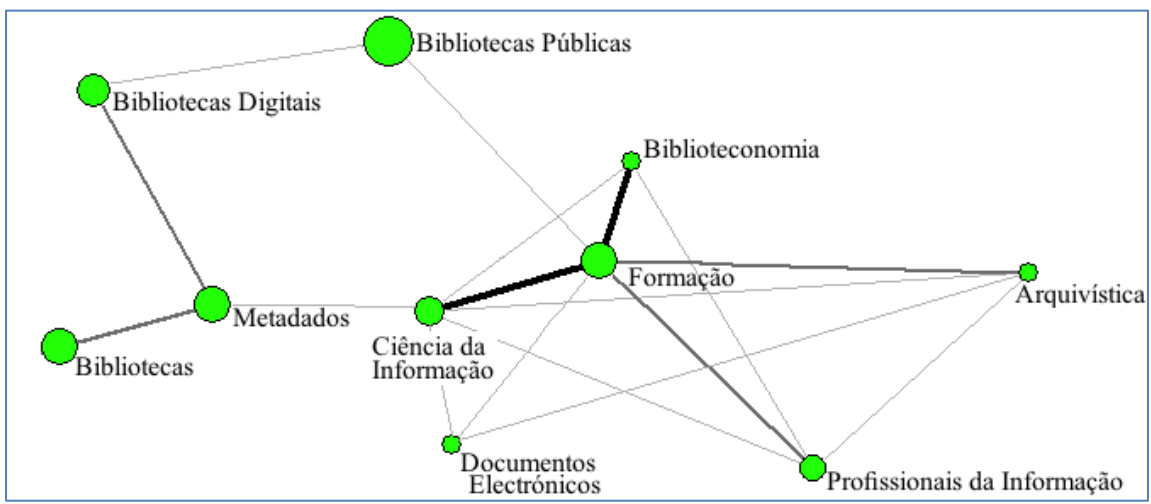

**Grafo 3.** Co-ocorrência de palavras-chave (Congressos 2001-2006). Pajek 3.11

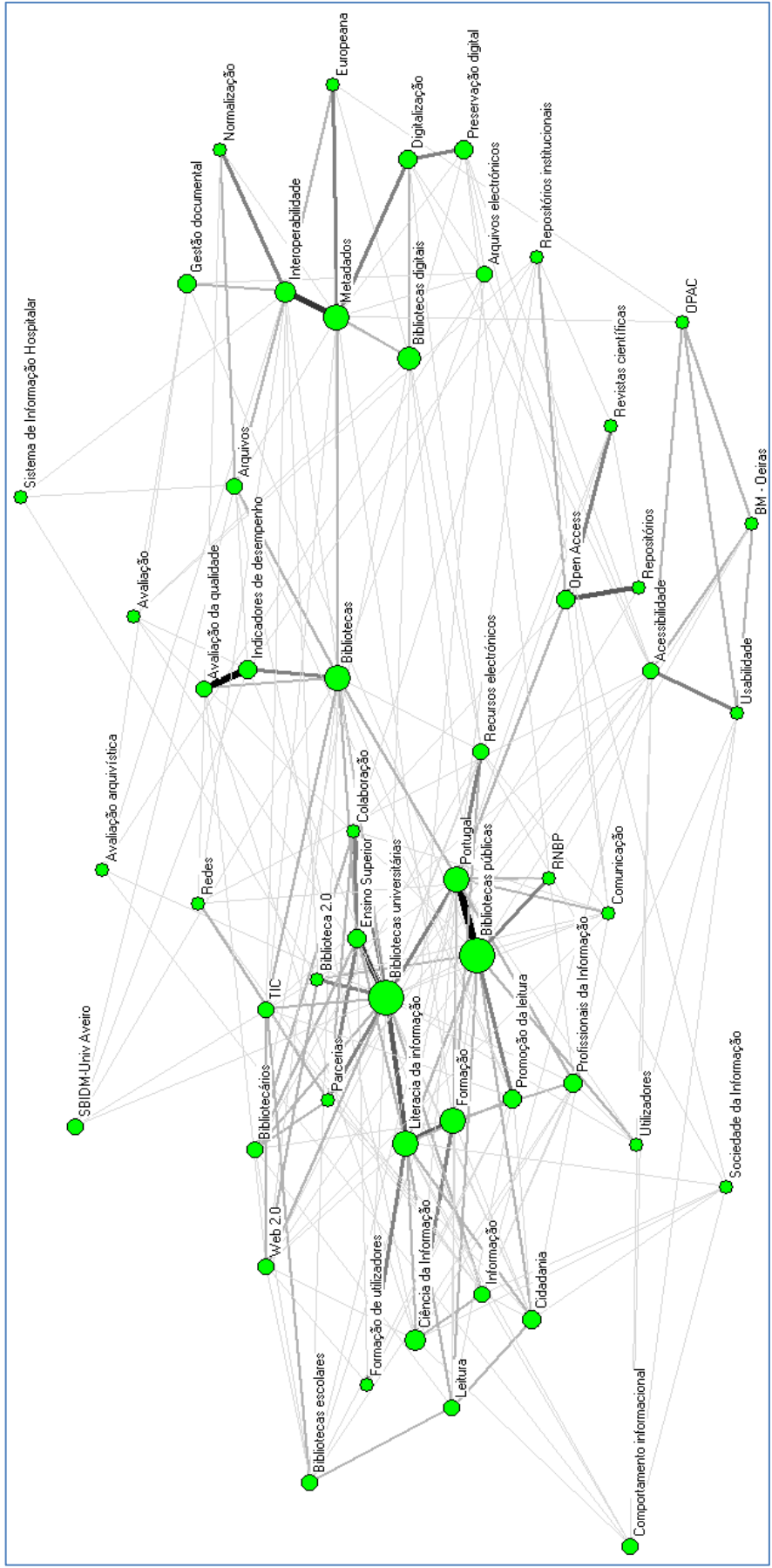

**Grafo 4.** Co-ocorrência de palavras-chave (Congressos 2007-2012) –Pajek 3.11

**Conclusões**

Apesar da sua periodicidade irregular, o Congresso chega a 2012 com uma participação superior ao valor médio de 64 comunicações por edição, reafirmando o seu lugar incontornável enquanto fórum dos profissionais da informação em Portugal.

Embora prevaleça a publicação individual e a colaboração, por norma, não ultrapasse o nível departamental, nota-se um incremento das redes de co-autoria, sendo no entanto ainda muito exígua a colaboração internacional. Genericamente, a base de participação institucional do Congresso é coerente com o seu escopo, tendo um perfil fundamentalmente nacional e profissional, ligado aos serviços públicos de informação, com destaque para o mundo das bibliotecas, tanto nas suas diferentes tipologias como nos quadros da Administração Pública, o que é corroborado pela abordagem temática.

A análise ao conteúdo das comunicações mostrou também como tem sido difícil ultrapassar o carácter emergente da disciplina em Portugal, com as temáticas ligadas à formação a manterem uma expressão muito significativa pela década de 2000 a dentro. De todas as formas, nota-se uma crescente presença do mundo académico entre os autores e uma capacidade geral para acompanhar os mais recentes desafios colocados pela evolução tecnológica e pelas novas formas de comunicação e relação social.

Finalmente, deve ter-se presente que os perfis aqui apresentados só têm validade no âmbito deste Congresso profissional, sendo necessário trabalho complementar para alargar este tipo de análises a outras tipologias documentais para um retrato mais profundo da produção técnica e científica na área da Ciência da Informação em Portugal.

**Referências bibliográficas**

**Silvana Roque de Oliveira**

Actualmente é assistente convidada na FCSH-UNL, no Curso de Mestrado em Ciências da Informação e da Documentação (CID), e aluna de doutoramento da Universidad de Alcalá, estando a preparar uma dissertação na área da Bibliometria, como bolseira de doutoramento da FCT. Desde este ano também é formadora na BAD.

É licenciada em História, tem mestrado em História da Expansão Portuguesa e o Curso de Especialização em CID, na FCSH-UNL, e obteve o DEA na Facultad de Documentación da Universidad de Alcalá.

Entre 2010 e 2012 foi coordenadora técnico-científica do projecto de tratamento documental da Biblioteca José Mattoso, na qualidade de representante do Curso de Mestrado em CID (FCSH-UNL), numa parceria entre o CAM e o IEM (FCSH-UNL).

Participou em vários projectos de investigação na área da História da Expansão Portuguesa e das CID. Também exerceu funções de assessoria no CHAM (FCSH-UNL).

Outra área de actuação tem sido a edição científica, sendo secretária do Conselho de Redacção da revista *Anais de História de Além-Mar* (indexada pela *Scopus*), desde 2011.

Actualmente está interessada no estudo dos padrões de produção e comunicação científicas nas Ciências Sociais e Humanas, bem como na história e teoria das CID em Portugal.

**Catarina Moreira**

Catarina Moreira é investigadora do Grupo de Sistemas de Informação do INESC-ID e é aluna de doutoramento em Engenharia Informática e de Computadores no Instituto Superior Técnico. No seu mestrado especializou-se em sistemas de informação empresariais e em sistemas inteligentes. A sua tese de mestrado consistiu na aplicação de algoritmos de aprendizagem automática para encontrar peritos em bases de dados de publicações académicas.

Em 2012, ganhou o Prémio da melhor dissertação de mestrado em Sistemas de Informação do Instituto Superior Técnico, e uma bolsa da Google para frequentar a *Lisbon Machine Learning Summer School*.

As áreas de investigação nas quais tem trabalhado são diversas, abrangendo estruturas de indexação de dados de elevadas dimensionalidades, algoritmos de aprendizagem automática, modelos probabilísticos de apoio à decisão e sistemas cognitivos quânticos.

**José Borbinha** é Professor do Departamento de Informática do IST e Investigador do Grupo de Sistemas de Informação do INESC-ID. É licenciado em Engenharia Eletrotécnica e de Computadores e doutorado em Engenharia Informática e de Computadores pelo IST.

Tem desenvolvido atividade de investigação em análise, modelação, desenho, integração e normalização em Sistemas de Informação, em cenários de bibliotecas, arquivos e "e-science".

Foi Diretor dos Serviços de Inovação e Desenvolvimento da Biblioteca Nacional de Portugal (1999 a 2004), onde lançou a Iniciativa da Biblioteca Nacional Digital e a renovação tecnológica da PORBASE.

Tem sido consultor da Comissão Europeia e NSF (National Science Foundation - Estados Unidos). Tem sido membro regular dos Comités de Programa das mais importantes conferências especializadas em bibliotecas digitais (JCDL, ECDL/TPDL, ICADL e RCDL), e membro do Comité Editorial do JDL – Journal of Digital Libraires.

É membro da Ordem dos Engenheiros (colégios de Engenharia Eletrotécnica e de Informática), IEEE - Institute of Electrical and Electronic Engineers, ACM - Association for Computer Machinery, INCOSE - International Council on Systems Engineering, e BAD - Associação Portuguesa de Bibliotecários, Arquivistas e Documentalistas. Foi membro fundador e "chair" eleito (2008 a 2010) do IEEE TCDL – Technical Committee on Digital Libraries.